\documentclass[a4paper,11pt]{article}
\usepackage{jheppub} 
\usepackage{lineno}
\usepackage{macros}
\usepackage{stmaryrd}
\usepackage{theorems}
\usepackage{amsthm}
\usepackage{mathtools}
\usepackage{dsfont}
\usepackage{cleveref}
\usepackage{dsfont}
\usepackage{amsmath,amssymb}
\usepackage{graphicx,multirow}
\usepackage{subcaption}
\usepackage[font=small,labelfont=bf]{caption}
\usepackage[compat=1.1.0]{tikz-feynman}
\usepackage{tikz}
\usetikzlibrary{decorations.pathmorphing}
\usepackage{caption}
\tikzset{snake it/.style={decorate, decoration=snake}}
\usetikzlibrary{shapes.geometric, arrows}
\usepackage{mathtools}
\usepackage{tabularx}
\usepackage{xcolor}
\usepackage{slashed}
\usepackage{units}

\def \L {\mathcal{L}} 
\newcommand{\hc}{\ensuremath{\text{h.c.}}}

\newcommand{\bea}{\begin{eqnarray}}
\newcommand{\eea}{\end{eqnarray}}
\newcommand{\beq}{\begin{eqnarray}}
\newcommand{\eeq}{\end{eqnarray}}

\renewcommand{\s}[1]{\slashed{#1}} 
\newcommand{\busp}[2]{\bar{u}^{#1}(#2)} 

\newcommand{\usp}[2]{u^{#1}(#2)}

\newcommand{\ii}{\mathrm{i}} 
\newcommand{\dd}{\mathrm{d}} 
\newcommand{\MM}{\mathcal{M}} 
\newcommand{\Group}[2]{\mathrm{#1(#2)}} 
\newcommand{\A}{\mathcal{A}} 
\renewcommand{\abs}[1]{\left| #1 \right|} 
\newcommand{\suL}{$\mathrm{SU(2)}_{\mathrm{L}}\ $}
\newcommand{\uY}{$\mathrm{U(1)}_{\mathrm{Y}}\ $}

\renewcommand{\im}[0]{\mathrm{Im}}
\newcommand{\eps}[0]{\epsilon}

\newcommand{\nx}[0]{n_{\chi_1}}
\newcommand{\Nx}[0]{N_{\chi_1}}
\newcommand{\vev}[1]{\langle #1 \rangle}

\allowdisplaybreaks

\tikzfeynmanset{warn luatex=false}

\definecolor{CBBlack}{RGB}{187,187,187}
\definecolor{CBOrange}{RGB}{230,159,0}
\definecolor{CBSkyBlue}{RGB}{86,180,233}
\definecolor{CBGreen}{RGB}{0,158,115}
\definecolor{CBYellow}{RGB}{240,228,66}
\definecolor{CBBlue}{RGB}{0,144,178}
\definecolor{CBBrown}{RGB}{213,94,0}
\definecolor{CBPurple}{RGB}{204,121,167}

\preprint{UCI-TR-2026-05}

\title{A Model of Annihilogenesis}

\author[a]{Arvind Rajaraman,}
\author[a]{Alexander Stewart,}
\author[a]{Tim M.P. Tait}

\emailAdd{arajaram@uci.edu}
\emailAdd{ajstewa1@uci.edu} 
\emailAdd{ttait@uci.edu}

\affiliation[a]{Department of Physics and Astronomy, University of California, Irvine\\
CA 92697-4575, USA}

\abstract{
We present an explicit model of leptogenesis via annihilogenesis in which two right-handed Majorana neutrinos couple to the Standard Model lepton doublets and Higgs, and acquire a large mass shift during a strong first-order phase transition of an additional scalar singlet. As bubbles of true vacuum expand, the $\chi_a$ are reflected off the walls and confined to shrinking pockets of false vacuum, where the density grows and the dominant CP-violating process is the $2 \to 4$ annihilation $\chi_1 \chi_1 \to L_1 L_1 \Phi^* \Phi^*$. Interference between tree-level $W$ and $B$ exchange and one-loop diagrams containing the heavier $\chi_2$ produces a CP asymmetry $\epsilon$, which we evaluate numerically and find to lie in the range $|\epsilon| \sim 10^{-9}$--$10^{-7}$ for $\mathcal{O}(1)$ Yukawa couplings. Electroweak sphalerons convert the resulting lepton asymmetry into a baryon asymmetry $Y_{\Delta B}$ that reproduces the observed value across a broad region of parameter space, with little sensitivity to the bubble-wall velocity or initial pocket size. The Majorana mass that controls $\epsilon$ is the residual mass of $\chi_1$ inside the collapsing pockets rather than its post-transition value, so the usual relation between the singlet mass and the Standard Model active neutrino masses is relaxed.  As a result, the upper bound on $|\epsilon|$ from the largest light-neutrino mass that constrains standard thermal leptogenesis does not apply, and the lower limits on the right-handed-neutrino scale and the reheating temperature are relaxed.
}

\begin{document}


\maketitle
\flushbottom

\section{Introduction}
The observed matter-antimatter baryon asymmetry of the Universe cannot be explained by the Standard Model (SM) alone, and requires additional physics. The creation of this asymmetry in the early Universe, referred to as baryogenesis (for reviews, see e.g.~Refs.~\cite{Bodeker:2020ghk,Cline:2006ts,Morrissey:2012db}), has inspired a wide range of mechanisms --- from below the electroweak (EW) scale up to the scale of grand unified theories (GUTs). All successful models satisfy the three conditions tabulated by Sakharov~\cite{Sakharov:1967dj}: baryon-number-violating processes, C and CP violation, and a departure from thermal equilibrium. The SM in principle contains all of the necessary ingredients, but the amount of CP violation in the Cabibbo-Kobayashi-Maskawa (CKM) matrix is too small to account for the observed asymmetry, and the electroweak phase transition is not expected to produce a sufficient departure from equilibrium. Many models of physics beyond the Standard Model (BSM) introduce fields and dynamics that can realize successful baryogenesis.


One way to satisfy the third Sakharov condition is to modify the Higgs sector to allow for a strong first-order phase transition (FOPT). A common realization is to extend the SM with new scalars that produce a strong FOPT~\cite{Gil:2012ya,Carena:2012np,Profumo:2014opa,Kozaczuk:2014kva,Vaskonen:2016yiu,Dorsch:2016nrg,Chiang:2017nmu,Beniwal:2018hyi,Bruggisser:2018mrt,Athron:2019teq,Kainulainen:2019kyp}. After the FOPT, bubbles of true vacuum nucleate and expand outwards, filling the Universe~\cite{Linde:1980tt,Langer:1969bc}. As these bubble walls sweep through the hot plasma of the early Universe, they separate particles by their spins, producing an excess of left-handed particles (and their negative-helicity anti-particles) on one side of the wall relative to the other. In the regions of false vacuum, electroweak sphalerons are unsuppressed and wash out any baryon asymmetry produced earlier. In the true vacuum, however, electroweak symmetry breaking (EWSB) suppresses sphaleron processes, and a net baryon asymmetry can survive --- leading to potentially successful baryogenesis~\cite{Morrissey:2012db}.

Electroweak sphalerons open a complementary direction. Sphalerons do not preserve baryon number $B$ or lepton number $L$ individually, but do preserve $B-L$, so any lepton asymmetry produced can be converted into a baryon asymmetry. The prototypical realization introduces singlet neutrinos of the kind often invoked to generate the light neutrino masses via the seesaw mechanism~\cite{Minkowski:1977sc,Yanagida:1979as,Glashow:1979nm,Gell-Mann:1979vob,Mohapatra:1980yp}; this mechanism for baryogenesis is known as leptogenesis, and was originally explored in Ref.~\cite{Fukugita:1986hr}. In leptogenesis, the Yukawa couplings of the new singlet neutrinos provide CP violation. Provided the rate of these Yukawa interactions is slower than the expansion rate $H$ of the Universe at the time the lepton asymmetry is generated, leptogenesis models also depart from thermal equilibrium, satisfying all three of Sakharov's conditions.

A well-studied realization, ``thermal leptogenesis,'' invokes hierarchical heavy neutrino masses $M_1 \ll M_{i>1}$~\cite{Endoh:2003mz,Buchmuller:1999cu,Luty:1992un,Gherghetta:1993kn,Plumacher:1996kc,Plumacher:1997ru,Covi:1996wh,Buchmuller:2000as,Buchmuller:2004nz,Giudice:2003jh,Pilaftsis:2005rv,Abada:2006ea,Nardi:2006fx,Abada:2006fw,Barbieri:1999ma,Chen:2007fv,Davidson:2007xu,Nardi:2007cf,Nardi:2007fs,Strumia:2006db,Strumia:2006qk}. In these models the heavy singlet neutrino $N_1$ is produced by scattering in the thermal bath, which allows the $N_1$ number density to be computed in terms of the seesaw parameters and the reheating temperature. With hierarchical masses, there is a lower bound on $M_1$ and consequently on the reheating temperature~\cite{Hamaguchi:2001gw,Davidson:2002qv}. This lower bound can come into tension with upper bounds on the reheating temperature in other models --- for example, supergravity scenarios that overproduce gravitinos at high reheating temperatures~\cite{Weinberg:1982zq,Khlopov:1984pf,Kawasaki:2008qe,Eberl:2024pxr,Kaneta:2023uwi}. The bound on $M_1$ is relaxed in cases where the heavy neutrino masses are not hierarchical but nearly degenerate, in which case CP violation can be resonantly enhanced by $N_i$--$N_j$ mixing~\cite{Flanz:1994yx,Pilaftsis:1998pd,Flanz:1996fb,Pilaftsis:2003gt,Arakawa:2024bkv}.

Given the importance of the baryon asymmetry as evidence for physics beyond the SM, it is crucial to map out the landscape of theories that can successfully realize baryogenesis through a variety of mechanisms. Annihilogenesis~\cite{Arakawa2021} is one such scenario, in which a strong FOPT at temperature $T$ produces a large shift in the mass of a BSM particle $\chi$, generating an interplay between the decay $\chi \to {\rm SM}$ and the annihilation $\chi\chi \to {\rm SM}$ during the transition. In the original construction $\chi$ was taken to have zero tree-level mass $m_\chi$ at temperatures $T' > T$, a condition we relax here. After the phase transition where a complex scalar $s$ acquires vacuum expectation value (VEV) $\vev{s}$, the $\chi$ modes acquire a large mass inside the bubbles of true vacuum (where $\vev{s} \neq 0$), $M_\chi^{\mathrm{in}} = y\vev{s} + m_\chi$, with $m_\chi$ a residual tree-level mass in the false vacuum (where $\vev{s} = 0$). As bubbles of true vacuum nucleate, expand, and collide, they fill the Universe and leave behind small contracting ``pockets'' of unbroken phase~\cite{Asadi:2021pwo}. If $M_\chi^{\mathrm{in}}/T \gg 1$, the majority of $\chi$ particles are reflected off the bubble walls and remain in the false vacuum, eventually becoming trapped in the collapsing pockets and ``squeezed'' into smaller regions. If the $\chi$ lifetime is shorter than the pocket-collapse time, $\chi$ depletes via decays; as the pocket collapses, the $\chi$ density grows and the annihilation channel can also efficiently deplete $\chi$. Which process dominates depends on the decay width $\Gamma_\chi$, the annihilation cross section $\vev{\sigma v}$, and the pocket collapse rate. Interference between tree- and loop-level diagrams in the annihilation can produce CP violation large enough for successful baryogenesis. Previous work has examined baryogenesis in similar scenarios with relativistic bubble walls~\cite{Baldes:2021vyz,Azatov:2021irb}, but assumed that the reflection of particles off the bubble walls was negligible.

We present an explicit model realizing annihilogenesis, in which new couplings between $\chi$ and the SM Higgs field enhance the CP violation. Bubble nucleation from the strong FOPT provides the departure from thermal equilibrium and the electroweak sphalerons furnish the baryon-number violation, satisfying all three Sakharov conditions. The dominant contribution to the lepton asymmetry arises from a $2 \rightarrow 4$ process rather than the $1 \rightarrow 2$ decay considered in standard thermal leptogenesis. We find that successful leptogenesis can be achieved through this scattering rather than through decay, which modifies the relations between the heavy right-handed neutrino masses and the other parameters of the model. As a result, the bounds on the right-handed-neutrino scale and on the reheating temperature are relaxed relative to standard thermal leptogenesis.

The paper is organized as follows. Section~\ref{sec:review} provides an overview of the general annihilogenesis scenario, and Section~\ref{sec:model} introduces the Lagrangian for the $\chi$ whose annihilation is responsible for the lepton asymmetry. Sections~\ref{sec:tree} and~\ref{sec:loops} detail the tree- and loop-level processes whose interference produces the CP-violation parameter $\epsilon$, and Section~\ref{sec:num_methods} describes the numerical methods used to evaluate the relevant cross sections. Section~\ref{sec:results} presents the numerical results and the conversion of $\epsilon$ into an observed baryon asymmetry $Y_{\Delta B}$. Section~\ref{sec:conc} contains our conclusions and outlook. Further details on the Feynman rules of the model and on various computations are provided in the appendices.

\section{Review of Annihilogenesis}\label{sec:review}

The most basic scenario for annihilogenesis has a fermion $\chi$ coupling to a complex scalar $s$. The thermal potential for $s$ is assumed to take a form such that, at some temperature $T$ in the early Universe, $s$ undergoes a FOPT and triggers bubble nucleation. The properties of the FOPT are largely set by the potential $V(s)$, including thermal corrections.

Generically, the finite-temperature potential takes the form~\cite{Arnold:1992fb}
\begin{equation}
    V(s,T) = D(T^2 - T_0^2)|s|^2 - ET |s|^3 + \frac{\lambda(T)}{4}|s|^4\;,
\end{equation}
where $D$, $E$, and $\lambda(T)$ can be expressed in terms of tree-level potential parameters, thermal loop corrections, and zero-temperature one-loop corrections. These parameters determine the critical temperature $T_c$, the nucleation temperature $T_n$, and the strength of the FOPT. For a strong FOPT, one requires $
E/\lambda(T_c) \gg 1$. Given a specific model, $D$, $E$, and $\lambda$ are calculable, with $E$ typically generated by interactions of bosonic fields coupled to $s$ --- for example, if the VEV of $s$ breaks a gauge symmetry, loop diagrams involving the relevant gauge bosons generate the cubic term, and $E$ is fixed by those loops.
These details are well-studied and not central to the annihilogenesis mechanism itself; we assume
that a large $E$ can be generated,
allowing for a strong FOPT.

Once bubbles of true vacuum have begun to fill the Universe and collide, small collapsing pockets of false vacuum remain.
The majority of $\chi$ particles inside these pockets
are confined within them,
since their tree-level mass in the false vacuum is much smaller than their mass in the true vacuum.
The $\chi$ are therefore out of equilibrium inside the collapsing pockets, while the SM fields are not confined during this process. Interactions between the SM Higgs, the lepton doublet, and $\chi$ introduce new CP-violating parameters that contribute to the generation of a lepton asymmetry during the collapse. The asymmetries in the decay and annihilation processes, $\eps_D$ and $\eps_A$, are parameterized as
\begin{subequations}
\begin{equation}
    \eps_D \equiv \frac{\sum_\alpha \big[ \Gamma(\chi \rightarrow \mathrm{SM}) - \Gamma(\chi \rightarrow \overline{\mathrm{SM}})]}{\sum_\alpha \big[ \Gamma(\chi \rightarrow \mathrm{SM}) + \Gamma(\chi \rightarrow \overline{\mathrm{SM}})]}\;,
\end{equation}
\begin{equation}
    \eps_A \equiv \frac{\sum_\alpha \big[ \sigma(\chi\chi \rightarrow \mathrm{SM\; SM}) - \sigma(\chi\chi \rightarrow \mathrm{\overline{SM}\; \overline{SM}})]}{\sum_\alpha \big[ \sigma(\chi\chi \rightarrow \mathrm{SM\; SM}) + \sigma(\chi\chi \rightarrow \mathrm{\overline{SM}\; \overline{SM}})]}\;,
\end{equation}
\end{subequations}
and together they control the final asymmetry generated in the lepton and baryon sectors.

\subsection{A Model of Annihilogenesis}\label{sec:model}

To realize a specific model of annihilogenesis, we introduce two right-handed Majorana fermions $\chi_a$ ($a=1,2$) that are gauge-singlets under the SM and couple to the left-handed leptons and the Higgs doublet (in addition to the couplings to the complex scalar $s$ that is responsible for the generation of the mass $M_\chi^{in}$, i.e.,~the mass inside the bubbles of true vacuum). The additional terms added to the SM Lagrangian are
\beq\label{eq:chi_lag}
	\L_{\chi}= \sum_{a=1,2}\ii\bar{\chi}_a\s{\partial}\chi_a + \sum_{\substack{a=1,2,\\B=1,2,3}}A_{Ba}\bar{\chi}_a \left( P_{L} L^{j}_{B} \right)\Phi^{i}\epsilon_{ij} + \frac{1}{2} \sum_{a,b=1,2}\overline{\chi^c}_a m_{ab} \chi_b + \hc\;,
\eeq
where $a=1,2$ labels the two $\chi_a$ particles, $B=1,2,3$ labels the SM generations, and $i,j$ label \suL fundamental indices. We work in a basis where the Majorana mass term for $\chi_a$ is diagonal, $m_{ab} = \mathrm{diag}(m_{\chi_1},m_{\chi_2})$ with $m_{\chi_1} < m_{\chi_2}$. 

We assume the phase transition that generates the heavy $\chi$ mass $M_\chi^{in}$ occurs before EWSB, so all leptons as well as the \suL and \uY gauge bosons are massless. As a result, the gauge interaction and mass eigenstates coincide. As described in Appendix~\ref{app:phases}, we use flavor rotations to eliminate all complex phases in the $A_{Ba}$ except for $A_{12}$. We further enforce $A_{31} = A_{32} = 0$, $A_{11} \in \mathbb{R}$, and $A_{22} \in \mathbb{R}$. Then $\chi_1$ only couples to $L_1$, and $\chi_2$ couples to both $L_1$ and $L_2$ but only the coupling to $L_1$ is complex.
The Feynman rules
arising from the new interactions added to the SM, as well as the Feynman rules for the relevant SM interactions (pre-EWSB), are collected in Appendix~\ref{app:feyn}.


Since $m_{\chi_1} < m_{\chi_2}$, we focus on the annihilation of the lighter species $\chi_1$, fixing the initial state of interest to $\chi_1 \chi_1$. The most obvious candidate, $\chi_1 \chi_1 \rightarrow L_B L_C$, violates hypercharge conservation. The process $\chi_1 \chi_1 \rightarrow L_B \overline{L}_C$ conserves hypercharge but produces no lepton asymmetry. Likewise, no $2 \rightarrow 3$ process produces a significant lepton asymmetry while conserving all gauge charges. The relevant process is $\chi_1 \chi_1 \rightarrow L_1 L_1 \Phi^* \Phi^*$, whose leading order Feynman diagram is shown in Eq.~(\ref{tikz:tree}). Since $\chi_1$ only couples to $L_1$, and the gauge and mass eigenstates coincide pre-EWSB, the only possible $L_B L_C$ combination in the final state is $L_1 L_1$.


\section{Tree-Level Annihilation}\label{sec:tree}

\begin{figure}[t]
	\centering
	\includegraphics[width=0.9\textwidth]{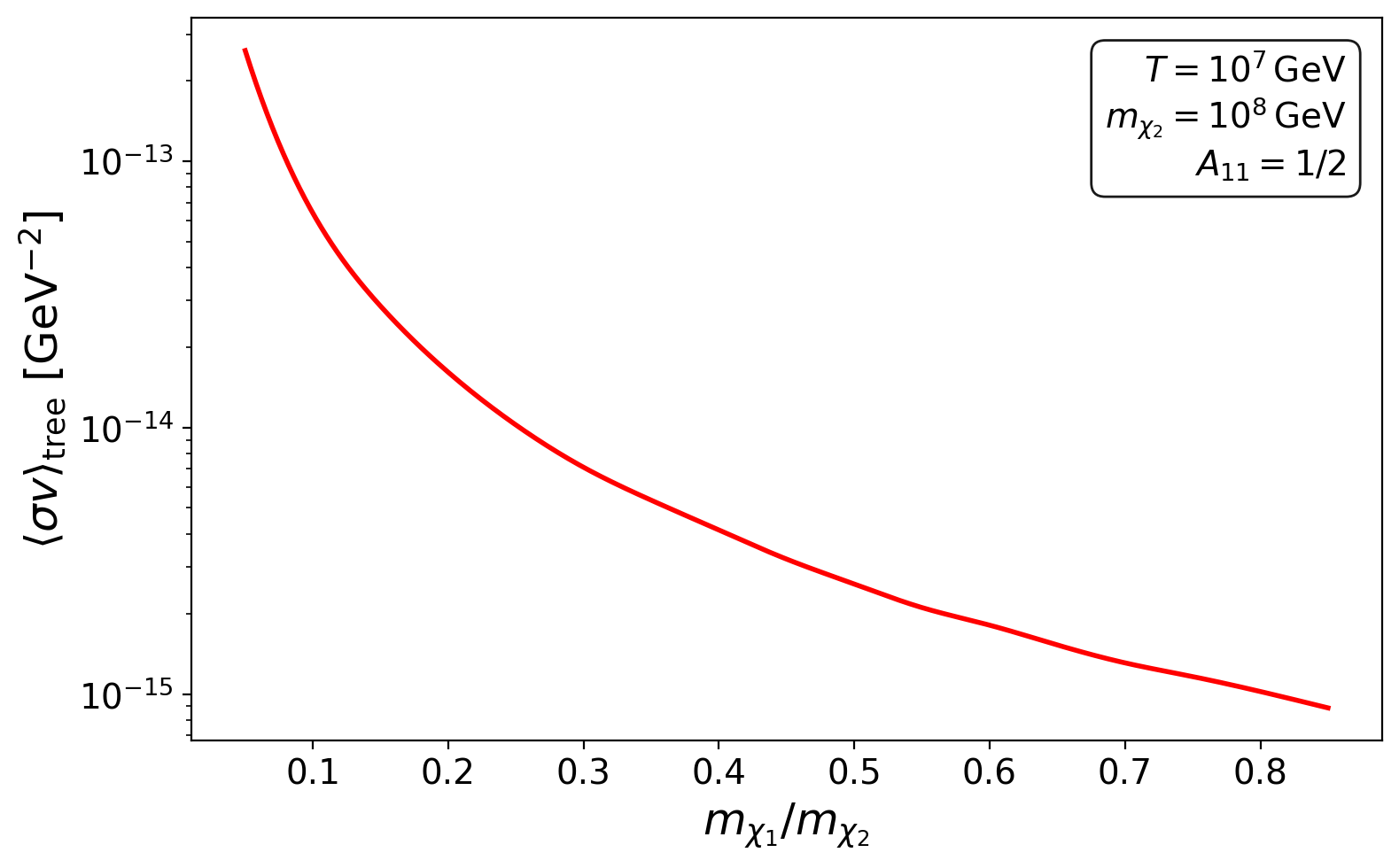}
	\caption{Tree level cross section $\langle \sigma v\rangle_{\mathrm{tree}}$ in GeV$^{-2}$ for the annihilation of two $\chi_1$ particles at rest as a function of the mass ratio $m_{\chi_1}/m_{\chi_2}$, for parameters: $T = 10^7$ GeV, $m_{\chi_2} =  10^8 $ GeV, and $A_{11} = 1/2$.}
	\label{fig:sigma_tree}
\end{figure}

We compute the relevant tree-level diagrams for $\chi_1 \chi_1 \rightarrow \Phi^* \Phi^* L_1 L_1$.
Since the Higgs Yukawas are smaller than the gauge couplings, the leading-order tree-level process is via $W$ (or $B$) exchange,
\beq\label{tikz:tree}
\begin{tikzpicture} [baseline=(i1.base)]
	\begin{feynman}
	\vertex (i1) {$\chi_1$};
	\vertex [below=of i1] (i2) {$\chi_1$};
	\vertex [right=of i1] (ri1);
	\vertex [right=of ri1] (rri1);
	\vertex [above=of rri1] (h1) {$\Phi^{*m}$};
	\vertex [right=of rri1] (o1) {$L_1^{o}$};
	\vertex [right=of i2] (ri2);
	\vertex [right=of ri2] (rri2);
	\vertex [below=of rri2] (h2) {$\Phi^{*n}$};
	\vertex [right=of rri2] (o2) {$L_1^{p}$};
		
	\diagram*{
		(i1) -- [solid] (ri1) -- [fermion, edge label'={$L_1^{j}$}] (rri1) -- [fermion] (o1),
		(i2) -- [solid] (ri2) -- [fermion, edge label={$L_1^{i}$} ] (rri2) -- [fermion] (o2),
		(ri1) -- [anti charged scalar] (h1),
		(ri2) -- [anti charged scalar] (h2),
		(rri1) -- [photon, edge label={$W^A$}] (rri2),
		};
		
	\end{feynman}
\end{tikzpicture}
\eeq
The resulting matrix element can be split into an amplitude $\A_{0}$ and coefficient $c_0$, where
\begin{eqnarray}\label{eq:treeAmpCoeff}
	\A_0 &=& g^{\mu\nu}D_W(p_1-q_1-q_3)\left[ \busp{v}{q_3}\gamma_\mu P_L S_{L_1}(p_1-q_1)P_R \usp{s}{p_1} \right] \nonumber \\
	&& \cdot \left[ \busp{w}{q_4}\gamma_\nu P_L S_{L_1}(p_2-q_2)P_R \usp{t}{p_2}\right]~, \\
	c_0 &=& A_{11}^2 g^2 t_{oj}^B t_{pi}^B \epsilon^{mj}\epsilon^{ni}~,
\end{eqnarray}
where the various quantities are defined in Appendix~\ref{app:feyn}.
The tree-level process and its coefficient are strictly real.
Numerical results for the thermally averaged cross section $\langle \sigma v\rangle$ (at threshold where $v \approx 0$) are presented in Fig.~\ref{fig:sigma_tree}. 

\section{Asymmetry at One Loop}\label{sec:loops}

We parameterize CP-violating effects in the annihilation of two $\chi_1$ particles through the parameter $\epsilon$, defined as
\beq\label{eq:eps}
	\epsilon \equiv \frac{\Gamma(\chi_1 \chi_1\rightarrow \Phi^* \Phi^* L_1 L_1)-\Gamma(\chi_1 \chi_1 \rightarrow \Phi \Phi \overline{L}_1 \overline{L}_1)}{\Gamma(\chi_1 \chi_1\rightarrow \Phi^* \Phi^* L_1 L_1)+\Gamma(\chi_1 \chi_1 \rightarrow \Phi \Phi \overline{L}_1 \overline{L}_1)}\;.
\eeq
CP asymmetry arises in the interference between tree- and loop-level processes. We split the tree- and loop-level matrix elements into a coupling constant $c$ and amplitude $\A$~\cite{Davidson:2008bu},
\beq
	\MM = \MM_0 + \MM_1 = c_0 \A_0 + c_1 \A_1\;,
\eeq
where $\MM_0$ is the tree-level matrix element and $\MM_1$ is the loop-level matrix element. The CP conjugate process has matrix element
\beq
	\overline{\MM} = c_0^* \overline{\A}_0 + c_1^* \overline{\A}_1\;.
\eeq
Using the fact that $\abs{\overline{\A}_i}^2 = \abs{\A_i}^2$, one finds that the CP asymmetry parameter $\epsilon$ has the form
\begin{eqnarray}\label{eq:eps_integral}
	\epsilon &=& \frac{\int\abs{c_0 \A_0 + c_1 \A_1}^2 \ \tilde{\delta}\ \dd\Pi_{11'} - \int \abs{c_0^* \A_0 + c_1^* \A_1}^2 \ \tilde{\delta}\ \dd\Pi_{11'}}{2 \int \abs{c_0 \A_0}^2 \ \tilde{\delta} \ \dd \Pi_{11'}} \nonumber \\
	&=& \frac{\im(c_0 c_1^*)}{\abs{c_0}^2}\frac{2\int \im(\A_0 \A_1^*) \ \tilde{\delta} \ \dd\Pi_{11'}}{\int \abs{\A_0}^2 \ \tilde{\delta} \ \dd\Pi_{11'}}\;,
\end{eqnarray}
where $\tilde{\delta}$ is the momentum conserving delta function,
\beq
	\tilde{\delta} \equiv (2\pi)^4 \delta^4 (P_i - P_f)\;,
\eeq
with $P_i$ and $P_f$ the summed initial and final momentum respectively, and $\dd\Pi_{11'}$ is the Lorentz invariant phase space (LIPS) measure,
\beq\label{eq:LIPS}
	\dd\Pi_{11'} \equiv \frac{\dd^3 p_{\Phi^*}}{2E_{\Phi^*}(2\pi)^3}\frac{\dd^3 p_{\Phi^{*'}}}{2E_{\Phi^{*'}}(2\pi)^3}\frac{\dd^3 p_{L_1}}{2E_{L_1}(2\pi)^3}\frac{\dd^3 p_{L_1'}}{2E_{L_1'}(2\pi)^3}\;.
\eeq

To compute the interference of the tree- and loop-level diagrams, we need the dispersive part of the loop diagrams, $\im(\A_0 \A_1^*)$. We obtain it using the Cutkosky cutting rules as described in Sec.~24.1.2 of Ref.~\cite{Schwartz:2014sze}. In the relevant loop diagrams the cuts occur on fermion propagators, where we replace
\beq
	S_j(p) = \frac{\s{p}+m_j}{p^2 - m_j^2 + i\eps} \rightarrow (\s{p}+m_j)\big[ -2\pi \ii \, \delta \big( p^2 - m_j^2 \big) \Theta( p^0 ) \big]\;.
\eeq

In addition, when $m_{\chi_1} > m_{\chi_2}/2$, the $\chi_2$ that appears in the loop diagrams can go on-shell. The pole arising from $\chi_2$ going on-shell is regulated using the standard Breit-Wigner propagator. One might worry that the cutting-rules computation of the imaginary part of the matrix element fails, since the Breit-Wigner propagator introduces an additional complex parameter. However, the propagator can be separated into its real and imaginary parts as,
\begin{equation}
	\frac{1}{p^2- m^2 + 2\ii \Gamma} = \frac{p^2 - m^2}{(p^2-m^2)^2 + m^2 \Gamma^2} - \frac{\ii m \Gamma}{(p^2-m^2)^2 - m^2 \Gamma^2}\;,
\end{equation}
so that, provided $\Gamma/m \ll 1$, the additional contribution to the imaginary part of the matrix element from $\chi_2$ going on-shell is negligible compared to the contribution from the cut diagrams considered above. Indeed, one finds
\begin{equation}
	\Gamma_{\chi_2} = \frac{m_{\chi_2}}{8\pi}\bigg(\sum_i A_{2i}\bigg)^2\;,
\end{equation}
so that, provided the couplings $A_{Ba}$ are not large, the additional imaginary piece arising from the Breit-Wigner propagator for $\chi_2$ can be neglected.

The tree-level process has no complex phases, so a complex phase (e.g., from $A_{12}$) is required in the diagram to produce a CP-violating process. The leading-order contribution arises from diagrams with both a Higgs and a $W$ (or $B$) gauge boson exchange. We present one representative example in detail; the remaining three loop diagrams are collected in Appendix~\ref{app:loops}.

An example relevant diagrams with its respective cut (shown as a dashed red line) is: 
\beq\label{img:L1}
\begin{tikzpicture} [baseline=(m1.base),scale=0.9,transform shape]
	\begin{feynman}
	\vertex (i1) {$\chi_1$};
	\vertex [below=of i1] (m1);
	\vertex [right=of m1] (m2);
	\vertex [right=of m2] (m3);
	\vertex [right=of m3] (m4);
	\vertex [right=of m4] (m5);
	\vertex [right=of m5] (m6);
	\vertex [right=of m6] (m7);
	\vertex [below=of m1] (i2) {$\chi_1$};
	\vertex [right=of i1] (u1);
	\vertex [right=of u1] (u2) {$\rho$};
	\vertex [right=of u2] (u3);
	\vertex [right=of u3] (u4);
	\vertex [right=of u4] (u5);
	\vertex [right=of u5] (o1) {$L_1^{o}$};
	\vertex [above=of u4] (h1) {$\Phi^{*m}$};
	\vertex [above=of u5] (h2) {$\Phi^{*n}$};
	\vertex [right=of i2] (b1);
	\vertex [right=of b1] (b2) {$\sigma$};
	\vertex [right=of b2] (b3);
	\vertex [right=of b3] (b4);
	\vertex [right=of b4] (b5);
	\vertex [right=of b5] (o2) {$L_1^{p}$};
		
	\diagram* {
		(i1) -- [solid] (u1) -- [anti fermion, edge label={$\overline{L}_1^{i}$}] (u2) -- [anti fermion, edge label={$\overline{L}_1^{\ell}$}] (u3) -- [solid, edge label'={$\chi_2$}] (u4) -- [fermion] (o1),
		(u3) -- [anti charged scalar] (h1),
		(u4) -- [anti charged scalar] (h2),
		(i2) -- [solid] (b1) -- [fermion, edge label'={$L_1^{k}$}] (b2) -- [fermion] (o2),
		(u1) -- [charged scalar, edge label={$\Phi^{j}$}] (b1),
		(u2) -- [photon, edge label={$W^B$}] (b2),
			};
	\draw [dashed, thick, red] (2.7,1) -- (2.7,-4.5);
	\end{feynman}
\end{tikzpicture}
\eeq
The imaginary part of the conjugated amplitude $\im(\A_{L1}^*)$ and coefficient $\im (c_{L1}^*)$ read:
\begin{eqnarray*}
    \im (\A_{L1}^*) &=& \big[ \busp{v}{q_3} P_R S_{\chi_2}(q_2 + q_3) P_R S_{\overline{L}_1}(-q_1-q_2-q_3) \gamma_\rho P_R (-\s{p}_1 + \s{k})P_L \usp{s}{p_1}\big] \\
	&& \cdot\, \big[ \busp{w}{q_4} \gamma_\sigma P_L (\s{p}_2 + \s{k})P_R \usp{t}{p_2}\big] \cdot \big( D_{\Phi}(k) D_{\Phi}(q_4 - p_2 - k) g^{\rho\sigma} \big) \\
	&& \cdot\, \big[-2\pi \ii \, \delta\big( (p_1 - k)^2\big) \Theta \big( p_1^0 - k^0 \big) \big] \big[-2\pi \ii \, \delta\big( (p_2 + k)^2\big) \Theta \big( p_2^0 + k^0 \big) \big]\;,
\end{eqnarray*}
and
\begin{equation*}
    \im (c_{L1}^*) = -\frac{1}{2}A_{12}^2 |A_{11}|^2 g^2 t_{i\ell}^B t_{pk}^B \eps^{no}\eps^{m\ell} \eps_{ji} \eps^{jk}\;.
\end{equation*}

\subsection{Numerical Methods}\label{sec:num_methods}

The integral in Eq.~(\ref{eq:eps_integral}) is too complicated to be evaluated analytically; we evaluate it numerically using Monte Carlo methods. Sampling and numerical integration are done using the \texttt{VEGAS} algorithm~\cite{PeterLepage1978}, while final-state momenta are generated using the \texttt{RAMBO} algorithm~\cite{Kleiss1986}. \texttt{RAMBO} generates massless four-momenta uniformly distributed throughout phase space, so the integration measure for the final states is weighted by the total phase space volume, given for an $n$-body massless final state with center-of-momentum energy $w$ as:
\beq
	V_n(w) = \frac{(\pi/2)^{n-1}w^{2n-4}}{\Gamma(n) \Gamma(n-1)}\;,
\eeq
with $\Gamma(n)$ the Euler gamma function. For our case, $n=4$ and $w = 2m_{\chi_1}$, this reduces to
\beq
	V_4(2 m_{\chi_1}) = \frac{\pi^3 m_{\chi_1}^4}{6}\;.
\eeq
Sampling with \texttt{RAMBO} and multiplying the integrand by $V_4(2 m_{\chi_1})/(2\pi)^8$ completes the integration over the final state phase space parameters.\footnote{The additional $(2\pi)^8$ factor arises from the normalization choices of the phase space integration in \texttt{RAMBO} \cite{Kleiss1986}, which does not include the $(2\pi)^3$ factors for each final state momenta or the $(2\pi)^4$ from the momentum conserving delta function.}

The loop momentum $k$ is sampled by drawing two additional numbers $x_1$ and $x_2$ uniformly in $[0,1]$. Momentum conservation, combined with the delta functions arising from the cuts, fixes $k = (0,m_{\chi_1}\hat{n})$, with $\hat{n}$ a unit $3$-vector. Integration over $k$ is given by
\begin{equation}
 	-\pi^2 \int_0^{2\pi} \dd \theta \int_0^\pi \sin(\phi) f\big|_{k^0 = 0, |\vec{k}| = m_{\chi_1}}.
 \end{equation}
Converting to integration over the random numbers $x_1$ and $x_2$ yields
 \begin{equation}
 	-\pi^2(2\pi^2) \int_0^{1} \dd x_1 \int_0^1 \dd x_2 \sin(\pi x_2) f\big|_{k^0 = 0, |\vec{k}| = m_{\chi_1}}
\end{equation}
where the Jacobian factor is $2\pi^2$.

\subsection{CP-Violation Parameter $\eps$}

The CP-asymmetry parameter in Eq.~(\ref{eq:eps_integral}) can be written,
\begin{equation}
	\epsilon = \frac{\im(A_{12}^2)}{\sigma_{\mathrm{Tree}}}\bigg[ \frac{(3g^2 + g'^2)^2}{3g^4 + g'^4}(\sigma_1 - \sigma_2) + \frac{3g^4 - 6 g^2 g'^2 - g'^4}{3g^4 + g'^4}(\sigma_3 - \sigma_4) \bigg]\;,
\end{equation}
where $\sigma_i$ denotes the integral over phase space of the imaginary part of the interference between the tree-level amplitude and loop $L_i$, and $\sigma_{\mathrm{Tree}}$ denotes the integrated tree amplitude squared.

\begin{figure}[h]
	\centering
	\includegraphics[width=0.9\textwidth]{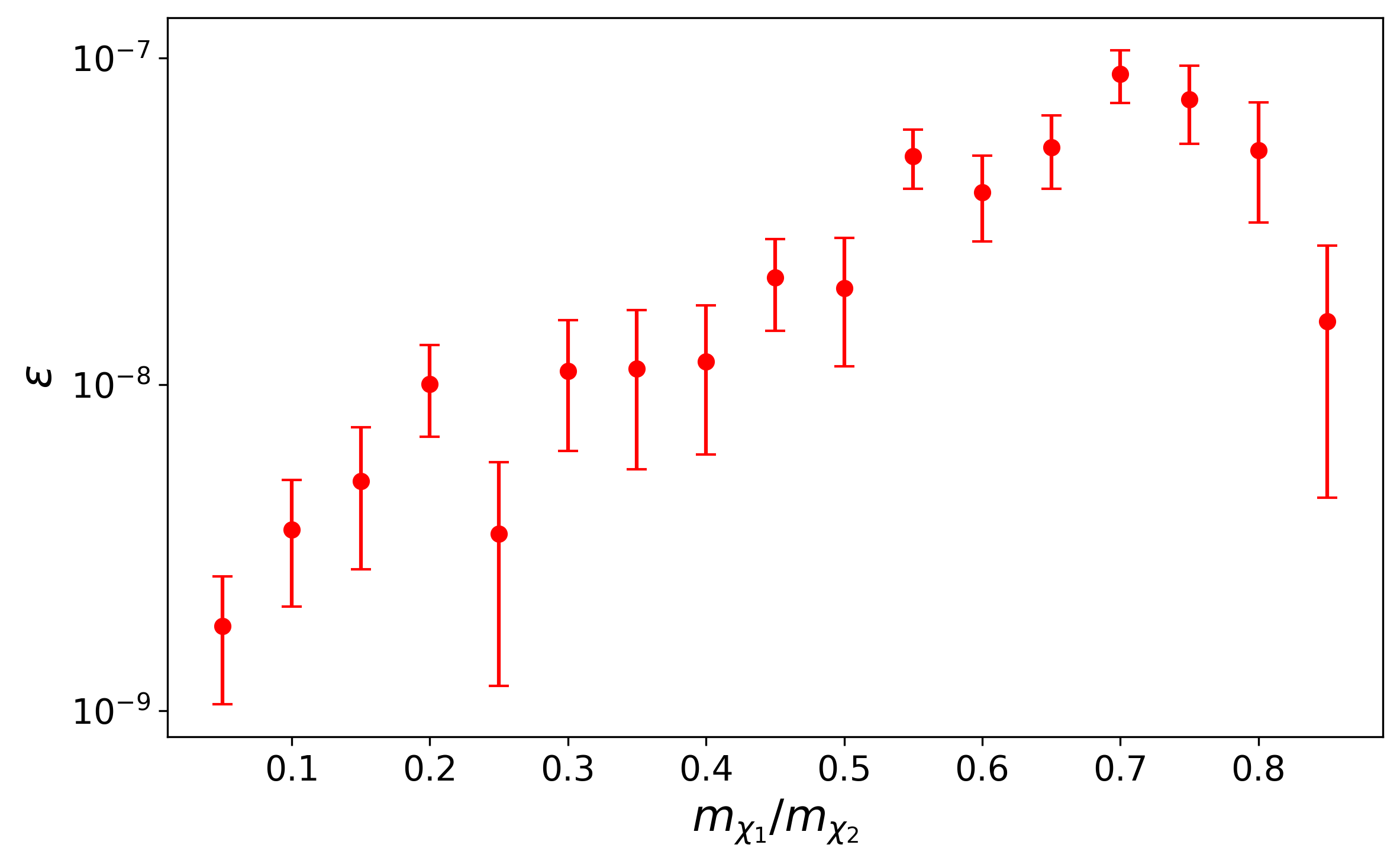}
	\caption{The CP-violation parameter $\eps$ as a function of $m_{\chi_1}/m_{\chi_2}$, for $g \approx 0.6$, $g' \approx 0.3$, $|A_{12}| =1$, and $\mathrm{arg}(A_{12}) = \pi/4$. Error bars indicate $1\sigma$ MC statistical uncertainties.}
	\label{fig:eps_plot}
\end{figure}

Our result for $\eps$ as a function of $m_{\chi_1} / m_{\chi_2}$ is shown in Fig.~\ref{fig:eps_plot}. The gauge couplings are taken near their $Z$-pole values, $g \approx 0.6$ and $g' \approx 0.3$, and we choose $|A_{12}| \approx \mathcal{O}(1)$ with $\mathrm{arg}(A_{12}) = \pi/4$. We find that the typical value of $\eps$ in this model lies in the range $10^{-9}$--$10^{-7}$, comfortably bracketing the value $\eps \sim 10^{-8}$ previously identified as being required for successful baryogenesis in the annihilogenesis scenario in Ref.~\cite{Arakawa2021}.

\section{Lepton and Baryon Asymmetries}\label{sec:results}

A lepton asymmetry is produced at scales well above the EWSB scale, which is converted into a baryon asymmetry by the unsuppressed electroweak sphalerons. Following Ref.~\cite{Arakawa2021}, to determine the total baryon asymmetry we compute the abundance of $\chi_1$ confined to collapsing pockets of unbroken phase, where the $\chi_a$ remain light. The Boltzmann equation for the number density of $\chi_1$, $\nx$, is
\begin{equation}
	\frac{\dd \nx}{\dd t} + 3\frac{\dot{R}}{R}\nx = -\langle \sigma v \rangle (\nx^2 - n_{\mathrm{eq}}^2) - \Gamma_{\chi_1}(\nx - n_{\mathrm{eq}})
\end{equation}
where $\Gamma_{\chi_1}$ is the decay width of the $\chi_1$ particle, $\langle \sigma v \rangle$ is the annihilation cross section for $\chi_1$, and $n_{\mathrm{eq}}$ is the equilibrium number density of the $\chi_1$ particles.

The initial sizes and bubble-wall velocities of the collapsing pockets of false vacuum are essential inputs. Defining $\beta_H = -T \tfrac{\d S}{\d T} |_{T_{\mathrm{nucl}}}$, one expects that the average number of bubbles that nucleate per Hubble volume during a strong FOPT scales as $N_b \sim \beta_H^3$, with $\beta_H$ typically of order $\mathcal{O}(10-10^3)$~\cite{Baldes:2021vyz}, though it may be as large as $\mathcal{O}(10^{11})$~\cite{Marfatia:2020bcs}. This fixes the initial size of pockets by specifying the number density of nucleating bubbles, $n_b \sim \beta_H^3 H^3$, and the distance between bubble centers, $d_b \sim n_b^{-1/3} \sim R_H / \beta_H$ with $R_H \equiv 1/H$~\cite{Megevand:2017vtb}. The speed at which the pockets collapse is set by the bubble wall velocity $v_w$, estimated as $v_w \sim (T_c - T)/T_c$~\cite{Witten:1984rs}. The pressure exerted on the bubble walls by the reflection of $\chi$ particles, however, can slow the bubble expansion considerably~\cite{Asadi:2021pwo,Baker:2021nyl,Marfatia:2021twj}.

In the original investigation of the annihilogenesis scenario~\cite{Arakawa2021}, both large and small initial radii $R_0 = R_H$ (corresponding to a few bubbles per Hubble volume) and $R_0 = 5\times 10^{-6} R_H$ (corresponding to many bubbles per Hubble volume) were studied as representative choices.  Both relativistic and non-relativistic cases were considered, with $v_w = 0.9$, $M_\chi^{\mathrm{in}}/T = 10^2$ and $v_w = 10^{-3}$, $M_\chi^{\mathrm{in}} = 10$ respectively, under the assumption that the wall velocity is approximately constant throughout the phase transition.

Assuming that the bubble wall moves at a constant velocity $v_w$, we recast the Boltzmann equation in terms of the radius of the bubble as
\begin{equation}
	\frac{\dd \nx}{\dd t} = \frac{\dd \nx}{\dd R}\dot{R} = -v_w \frac{\dd \nx}{\dd R}\;. 
\end{equation}
The total number of $\chi_1$ within the bubble is $\Nx = 4\pi R^3 \nx / 3$, so the Boltzmann equation can be converted into an equation for $\Nx$,
\begin{equation}
	\frac{\dd \Nx}{\dd R} = -\frac{4\pi R^3}{3 v_w} \bigg( \langle \sigma v \rangle (\nx^2 - n_{\mathrm{eq}}^2) + \Gamma_{\chi_1}(\nx - n_{\mathrm{eq}}) \bigg)
\end{equation}
The total asymmetry generated by the decay and annihilation of $\chi_1$ is,
\begin{equation}
	\epsilon_{\mathrm{tot}} = \eps_A f_A + \eps_D (1-f_A)\;,
\end{equation}
where $\eps_A$ and $\eps_D$ are the asymmetries from annihilations and decays, respectively, and $f_A$ is the fraction of the $\chi_1$ that annihilate versus decay. In the annihilogenesis scenario, confining the $\chi_1$ particles to collapsing pockets leads to $\langle \nx \rangle \gg 1$, so that $f_A \approx 1$ and $\eps_{\mathrm{tot}} \approx \eps_A$. The resulting baryon asymmetry is parameterized as,
\begin{equation}
	Y_{\Delta B} = \frac{\Delta \nx (T)}{s(T)}\eps_A C\;,
\end{equation}
where $\Delta \nx(T)$ is the total change in the number density of the $\chi_1$ during the pocket collapse, $s(T)$ is the entropy density, and $C$ is a constant that converts the CP asymmetry in the $\chi_1$ depletion into a baryon asymmetry. In our case, electroweak sphalerons convert the produced lepton asymmetry into the baryon asymmetry, and $C \simeq 12/37$~\cite{Davidson:2008bu}.

In the non-relativistic limit where $m_{\chi_1}/T \ll 1$, the equilibrium number density $n_{\m{eq}}$ is given by the standard Maxwell-Boltzmann distribution,
\begin{equation}
    n_{\m{eq}} = g\bigg( \frac{m_{\chi_1} T}{2 \pi} \bigg)^{3/2}e^{-m_{\chi_1}/T} \quad (m_{\chi_1}/T \ll 1)\;,
\end{equation}
where $g=2$ are the number of degrees of freedom of the Majorana fermion $\chi_1$. The entropy density $s(T)$ at temperature $T$ is given by
\begin{equation}
    s(T) = \frac{2\pi^2}{45}g_* T^3\;,
\end{equation}
where $g_* \approx 106.75$ counts the effective degrees of freedom of the SM in the early Universe.

\begin{figure}[t]
	\centering
	\includegraphics[width=0.495\textwidth]{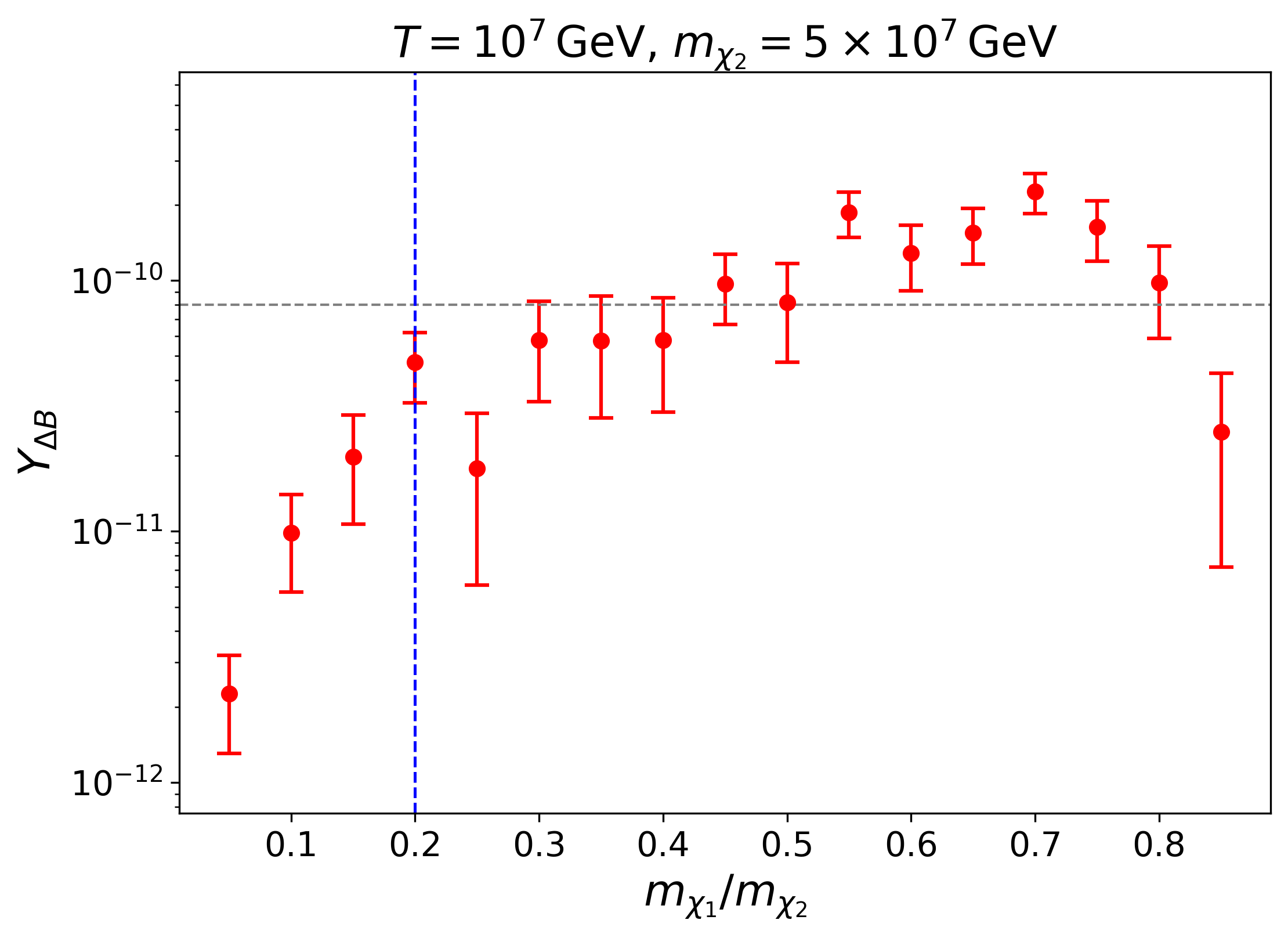}
    \includegraphics[width=0.495\textwidth]{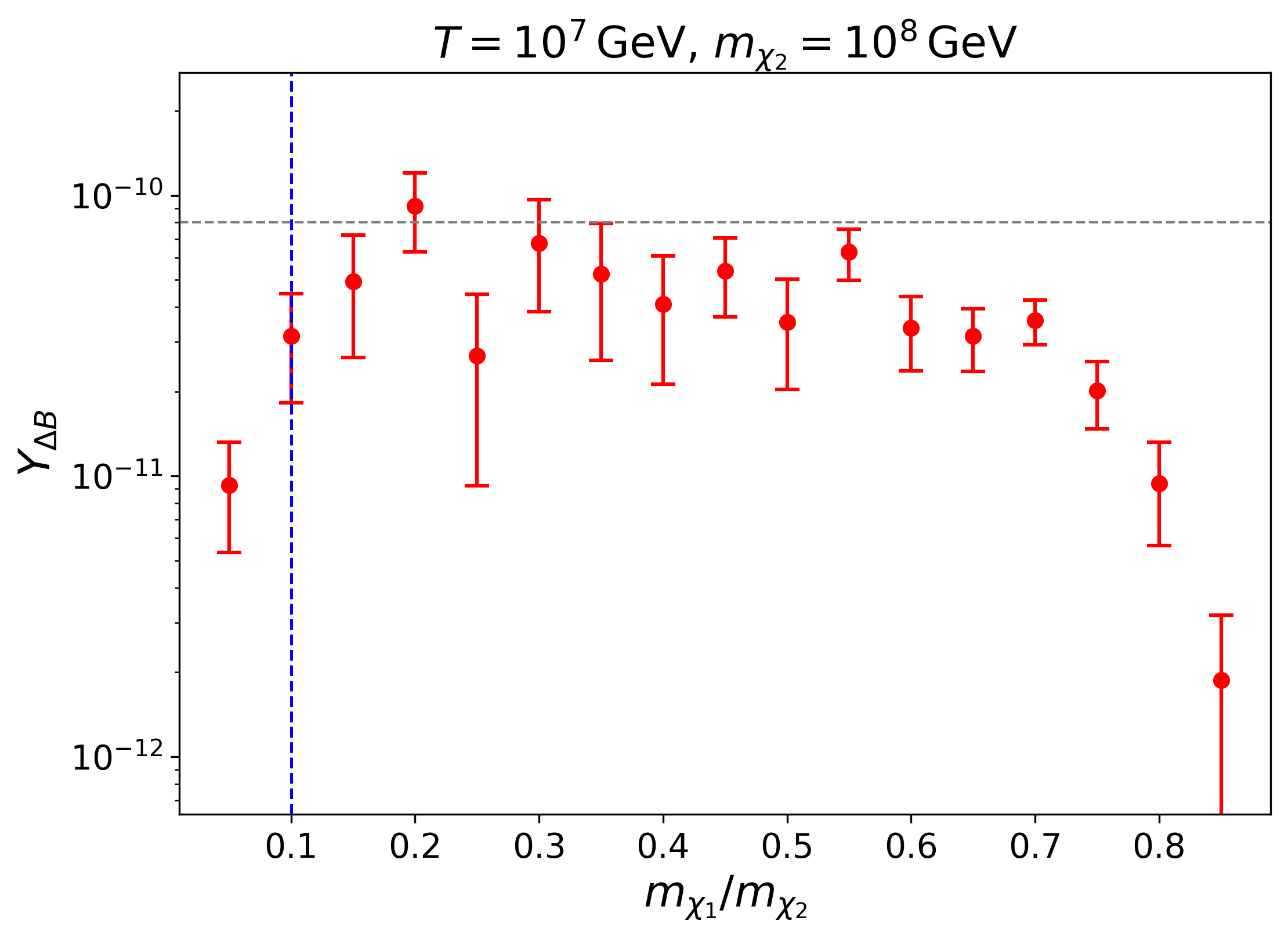}
	\caption{The computed baryon asymmetry $Y_{\Delta B}$ as a function of $m_{\chi_1}/m_{\chi_2}$, for $T=10^7$~GeV and $m_{\chi_2} = 5\times 10^7$~GeV (left panel) and $T=10^7$~GeV and $m_{\chi_2} = 10^8$~GeV (right panel). The dashed black line indicates the baryon asymmetry $Y_{\Delta B} \sim 8\times 10^{-11}$ required for successful baryogenesis. The vertical blue dashed line marks the point where $m_{\chi_1} = T$ and the non-relativistic approximation breaks down. Error bars indicate $1\sigma$ MC statistical uncertainties.}
	\label{fig:YB}
\end{figure}

Representative numerical results for the baryon asymmetry generated during pocket collapse are further shown in Fig.~\ref{fig:YB}. We find that when $m_{\chi_2}/T \lesssim \C{O}(10)$, the resulting depletion of $\chi_1$ is dominated by annihilation processes, with $f_A \approx 1$. The annihilation proceeds much faster than pocket collapse, so all $\chi_1$ are depleted by the end of the collapse and $Y_{\Delta B}$ is limited by the initial density of the pockets. The generated baryon asymmetry is therefore insensitive to the wall velocities $v_w$ and initial pocket sizes $R_0 / R_H$, and depends only on the ratios $m_{\chi_2}/T$ and $m_{\chi_1}/m_{\chi_2}$. We verify this numerically by checking that $Y_{\Delta B}$ for $(T,m_{\chi_2}) = (10^7,5\times 10^7)$~GeV is identical to the $(10^3,5\times 10^3)$~GeV case, as are the $(10^7,10^8)$~GeV and $(10^3,10^4)$~GeV cases. In these parameter regions $Y_{\Delta B}$ shows no sensitivity to $v_w$ in the range $(10^{-3},0.9)$ or to $R_0/R_H$ in the range $(5\times 10^{-6},1)$. Successful baryogenesis is achieved in both cases over a wide range of masses for ${\cal O}(1)$ CP-violating phases.

\subsection{Light Neutrino Masses}
In the type-I seesaw model, light neutrino masses are generated by the inclusion of heavy right-handed Majorana neutrinos with mass matrix $M_R$, which allow for neutrino Yukawa couplings. Once the Higgs acquires a VEV $v_h$, the neutrinos pick up a Dirac mass term encoded in the matrix $m_D$. The effective mass matrix for the light neutrinos then takes the form
\begin{equation}
    m_{\nu} = \big(m_D M_R^{-1} m_D^T\big)\;,
\end{equation}
where the heavy right-handed neutrinos naturally account for the small masses of the observed left-handed neutrinos through the $M^{-1}_R$ suppression of $m_\nu$. In the leptogenesis scenario in which the lepton asymmetry is generated by the decay of the lightest right-handed neutrino $N_1$, the CP-violation parameter $\eps$ and the largest light neutrino mass $m_{\mathrm{max}}$ can be related~\cite{Davidson:2008bu}. For three right-handed neutrinos with hierarchical masses $M_2, M_3 \gg M_1$, one can derive the upper bound
\begin{equation}
    |\eps| \leq \frac{3 M_1 m_{\mathrm{max}}}{16 \pi v_h^2} \sqrt{B^{N_1}_{\Phi L} + B^{N_1}_{\Phi^* \overline{L}}}\;,
\end{equation}
with $B^{N_1}_{xy}$ the branching ratio for the decay $N_1 \rightarrow xy$.

In the annihilogenesis scenario, we have shown that the CP-violation parameter $\eps$ depends instead on the ratio of the heavy right-handed Majorana masses, $m_{\chi_1} / m_{\chi_2}$. As the bubbles of false vacuum collapse, the confinement of $\chi_1$ to the interior of the bubble produces a pressure on the bubble wall that arrests the collapse until the $\chi_1$ have annihilated; the value of $m_{\chi_1}/m_{\chi_2}$ at this point fixes the value of $\eps$ relevant for leptogenesis. Once all bubbles of false vacuum have collapsed, the $\chi_a$ acquire the larger masses $M_{\chi}^{in}$ generated by the original phase transition. It is these $M_{\chi}^{in}$ that appear in the right-handed Majorana mass matrix $M_R$ relevant for the light neutrino masses. As a result, in the annihilogenesis model there is no relation between $\eps$ and $m_{\mathrm{max}}$ --- in contrast to the standard leptogenesis case, where the light neutrino masses constrain $|\eps|$ from above.

\section{Conclusions}\label{sec:conc}

We explore an explicit model of Annihilogenesis consisting of two generations of right-handed Majorana fermions $\chi_a$ with Yukawa couplings to the Higgs and lepton doublets of the SM. We find that the typical value of $|\eps|$ falls in the range $10^{-9}$--$10^{-7}$, sufficient for successful baryogenesis. In contrast to leptogenesis proceeding via the decay of heavy right-handed fermions, in annihilogenesis there is no strict upper bound on $|\eps|$ from the light-neutrino masses and the heavy right-handed-neutrino masses. As a consequence, the annihilogenesis realization of leptogenesis can evade the constraints relating $|\eps|$ to the light-neutrino masses while still producing the CP violation needed to generate the observed matter-antimatter asymmetry.

The mechanism explored here suggests several directions worthy of further study. The interplay between $\chi$ scattering during pocket collapse and the cosmological history of the FOPT itself deserves a more detailed treatment, in particular with respect to the back-reaction of the trapped $\chi$ on the bubble-wall dynamics and the resulting gravitational-wave signal. Extensions to non-minimal flavor structure, or to alternative scalar potentials, may further open the parameter space accessible to the mechanism. We look forward to seeing these questions addressed in future work.

\section*{Acknowledgments}
A.S. was supported by the National Science Foundation Graduate Research Fellowship Program. This material is based upon work supported by the National Science Foundation Graduate Research Fellowship Program under Grant No. DGE-2235784. Any opinions, findings, and conclusions or recommendations expressed in this material are those of the authors and do not necessarily reflect the views of the National Science Foundation.  The research of A.R. and T.M.P.T. was supported in part by the National Science Foundation through Grant PHY-2210283.  T.M.P.T. acknowledges additional NSF support via Grant PHY-2514888.

\appendix



\section{Complex Phases}\label{app:phases}

The Lagrangian contains the terms,
\begin{equation}
    \mathcal{L}_\chi \supset \sum_{B = 1,2,3}\bigg[\sum_{a=1,2} A_{Ba}\overline{\chi}_a \big(P_L L_B^j\big) \Phi^i \epsilon_{ij} + g \overline{L}_{i,B} \gamma^\mu (t^A)^i{}_j L_B^j W_\mu^A - \frac{g'}{2} \overline{L}_{i,B} \gamma^\mu L_B^i B_\mu \bigg]\;.
\end{equation}
As in \ref{eq:chi_lag}, index $a=1,2$ labels the $\chi_a$ generation, $B=1,2,3$ the SM generation, $i,j$ are $\Group{SU}{2}_{\mathrm{L}}$ fundamental indices, and here $A$ is the $\Group{SU}{2}_{\mathrm{L}}$ adjoint index with $(t^A)^i{}_j$ the generators of $\Group{SU}{2}_{\mathrm{L}}$.
From this Lagrangian, it is evident that $L_B$ has a $\mathrm{U}(3)_{L_B}$ flavor symmetry under which:
\begin{equation}
    L_B \xmapsto{V \in \mathrm{U}(3)_{L_B}} V L_B\;,\quad A_{Ba} \xmapsto{V \in \mathrm{U}(3)_{L_B}} A_{Ba}V^\dagger\;.
\end{equation}
Before EWSB, the leptons do not have any mass terms, and the mass and gauge bases can be chosen to be identical. Nonetheless, the Yukawa couplings between the Higgs doublet, left-handed lepton doublet, and right-handed lepton singlet break the symmetry, but these Yukawa interactions are much weaker than the $\mathrm{SU}(2)_\mathrm{L} \times \mathrm{U}(1)_\mathrm{Y}$ gauge couplings, and give subdominant contributions to the CP violation from annihilation of $\chi_1$. For this reason, the charged-lepton Yukawas can be neglected and do not appear in the loop diagrams we consider for $\eps$. The Yukawas between the $\chi_i$ and lepton doublet do appear, since such interactions are not necessarily negligible, and are necessary to generate CP violation.

The $\mathrm{U}(3)_{L_B}$ flavor symmetry allows one to rotate the $A_{1a}$ such that only $A_{11} \in \mathbb{R}$ is non-zero. This leaves a $\mathrm{U}(2)_{L_B}$ flavor symmetry that rotates $L_2$ and $L_3$. We exhaust this remaining $\mathrm{U}(2)_{L_B}$ symmetry by demanding that $A_{22} \in \mathbb{R}$ and $A_{23} = 0$, leaving one complex coupling $A_{21} \in \mathbb{C}$. In this basis the Lagrangian contains terms of the form
\begin{equation}
    \mathcal{L}_\chi \supset A_{11} \overline{\chi}_1 L_1 \Phi + A_{21} \overline{\chi}_2 L_1 \Phi + A_{22} \overline{\chi}_2 L_2 \Phi + \dots 
\end{equation}
where the ``$\dots$'' indicate the gauge interactions of the leptons which are unchanged under $\mathrm{U}(3)_{L_B}$ rotations. Note that $\chi_1$ only couples to $L_1$, while $\chi_2$ couples to both $L_1$ and $L_2$, but only the coupling to $L_1$ contains the complex coupling coefficient $A_{21}$.

\section{Feynman Rules}\label{app:feyn}

Arrows on lines in Feynman diagrams track hypercharge $Y=-1/2$, so arrows on Dirac fermions also provide a way to track fermion-number flow. Index $A$ denotes the $\Group{SU}{2}_\mathrm{L}$ adjoint, while $i$ and $j$ label $\Group{SU}{2}_\mathrm{L}$ fundamental and anti-fundamental indices. The index $a=1,2$ labels the two Majorana fermions $\chi_a$, and $B=1,2,3$ labels the lepton doublet generation $L_B$.

We use the rules of Ref.~\cite{Denner1992} for computing amplitudes involving Majorana fermions. The fermion-number flow is taken to be directed from left to right along both the upper and lower fermion lines in all diagrams (see Secs.~\ref{sec:tree}, \ref{sec:loops}, and Appendix~\ref{app:loops}). To accommodate Majorana fermions, the following additional rules are implemented:
\begin{enumerate}
	\item Dirac fermion propagators with fermion flow against the orientation are assigned propagators with the opposite momenta, i.e.,~$S(p) \rightarrow S(-p)$
	\item Vertices involving Dirac fermions are charge conjugated if the fermion flow at the vertex is against the orientation
\end{enumerate}
The two main effects of these additional rules are on the $\gamma$ matrices and on the $\Group{SU}{2}_\mathrm{L}$ fundamental and anti-fundamental indices. The relevant results for charge conjugation of the $\gamma$ matrices are $C \gamma_\mu^T C^{-1} = -\gamma_\mu$ and $C(\gamma_\mu \gamma_5)^T C^{-1} = -\gamma_\mu \gamma_5$. For $\Group{SU}{2}_\mathrm{L}$ fundamental and anti-fundamental fields, charge conjugation sends the representation to its conjugate representation: the $\Group{SU}{2}_{\mathrm{L}}$ generators go from $t^{A}_{ij}$ to $-t^A_{ji}$, and the $\epsilon_{ij}$ are transposed as well (for the $\epsilon_{ij}$, this just gives an overall minus sign once we re-order indices). The generators of $\Group{SU}{2}_\mathrm{L}$ are taken to have normalization $t_{ij}^A = (1/2) \sigma^A_{ij}$, where $\sigma^A_{ij}$ are the Pauli matrices. With these conventions, the relevant charge-conjugated vertices, assuming overall fermion-number orientation from left to right, are as follows. First, the new Feynman rules introduced to the SM by the coupling of $\chi_a$, lepton doublet $L_B$, and Higgs $\Phi$,

\begin{equation*}
\begin{array}{c@{\qquad}c@{\qquad}c}

\begin{gathered}
  \begin{tikzpicture} [baseline=(v.base),scale=0.9, transform shape]
	\begin{feynman}
	\vertex (v);
	\vertex [left=of v] (i1) {\(\chi_a\)};
	\vertex [right=of v] (o1) {\(L^{j}_B\)};
	\vertex [above=of o1] (o2) {\(\Phi^{*i}\)};
	
	\diagram* {
		(i1) -- [plain] (v) -- [fermion] (o1),
		(o2) -- [charged scalar] (v),
	};
	\end{feynman}
  \end{tikzpicture}
  \\[0.7ex]
  \ii A_{Ba}\,\epsilon^{ij} P_{R}
\end{gathered}

&
\begin{gathered}
 \begin{tikzpicture} [baseline=(v.base),scale=0.9, transform shape]
	\begin{feynman}
		\vertex (v);
		\vertex [left=of v] (i1) {\(\overline{L}^{j}_B\)};
		\vertex [right=of v] (o1) {\(\chi_a\)};
		\vertex [above=of o1] (o2) {\(\Phi^{*i}\)};
		
		\diagram* {
			(i1) -- [anti fermion] (v) -- [plain] (o1),
			(o2) -- [charged scalar] (v),
		};
	\end{feynman}
  \end{tikzpicture}
  \\[0.7ex]
  -\ii A_{Ba} \epsilon^{ij} P_{R}
\end{gathered}

&
\begin{gathered}
  \begin{tikzpicture} [baseline=(v.base),scale=0.9, transform shape]
	\begin{feynman}
		\vertex (v);
		\vertex [left=of v] (i1) {\(\chi_a\)};
		\vertex [right=of v] (o1) {\(\overline{L}^{j}_B\)};
		\vertex [above=of o1] (o2) {\(\Phi^{i}\)};
		
		\diagram* {
			(i1) -- [plain] (v) -- [anti fermion] (o1),
			(o2) -- [anti charged scalar] (v),
		};
	\end{feynman}
  \end{tikzpicture}
  \\[0.7ex]
  -\ii A_{Ba}^{*} \epsilon_{ij} P_{L}
\end{gathered}

\end{array}
\end{equation*}
Additionally, the following Feynman rules arise from the SM Lagrangian pre-EWSB,

\begin{equation*}
\begin{array}{c@{\qquad}c@{\qquad}c}

\begin{gathered}
  \begin{tikzpicture} [baseline=(v.base),scale=0.9,transform shape]
	\begin{feynman}
	\vertex (v);
	\vertex [left=of v] (i1) {\(L^{j}_{B}\)};
	\vertex [right=of v] (o1) {\(L^{i}_{B}\)};
	\vertex [above=of o1] (o2) {\(W^A\)};
	\vertex [below=0.25em of v] {\(\mu\)};
	
	\diagram* {
		(i1) -- [fermion] (v) -- [fermion] (o1),
		(v) -- [photon] (o2),
	};
	\end{feynman}
\end{tikzpicture}
  \\[0.7ex]
  \ii g t_{ij}^A \gamma_\mu P_L
\end{gathered}

&
\begin{gathered}
  \begin{tikzpicture} [baseline=(v.base),scale=0.9,transform shape]
	\begin{feynman}
		\vertex (v);
		\vertex [left=of v] (i1) {\(\overline{L}^{j}_{B}\)};
		\vertex [right=of v] (o1) {\(\overline{L}^{i}_{B}\)};
		\vertex [above=of o1] (o2) {\(W^A\)};
		\vertex [below=0.25em of v] {\(\mu\)};
		
		\diagram* {
			(i1) -- [anti fermion] (v) -- [anti fermion] (o1),
			(v) -- [photon] (o2),
		};
	\end{feynman}
\end{tikzpicture}
  \\[0.7ex]
  \ii g t_{ji}^A \gamma_\mu P_R
\end{gathered}

&
\begin{gathered}
 \begin{tikzpicture} [baseline=(v.base),scale=0.9,transform shape]
	\begin{feynman}
	\vertex (v);
	\vertex [left=of v] (i1) {\(\Phi^{j}\)};
	\vertex [right=of v] (o1) {\(\Phi^{i}\)};
	\vertex [above=of o1] (o2) {\(W^A\)};
	\vertex [below=0.25em of v] {\(\mu\)};
	
	\diagram* {
		(i1) -- [charged scalar,momentum'={$p$}] (v) -- [charged scalar,momentum'={$q$}] (o1),
		(v) -- [photon] (o2),
	};
	\end{feynman}
\end{tikzpicture}
  \\[0.7ex]
  -\ii g t_{ij}^A (p+q)_\mu
\end{gathered}

\end{array}
\end{equation*}

\begin{equation*}
\begin{array}{c@{\qquad}c@{\qquad}c}

\begin{gathered}
  \begin{tikzpicture} [baseline=(v.base),scale=0.9,transform shape]
	\begin{feynman}
	\vertex (v);
	\vertex [left=of v] (i1) {\(L^{i}_{B}\)};
	\vertex [right=of v] (o1) {\(L^{j}_{B}\)};
	\vertex [above=of o1] (o2) {\(B\)};
	\vertex [below=0.25em of v] {\(\mu\)};
	
	\diagram* {
		(i1) -- [fermion] (v) -- [fermion] (o1),
		(v) -- [photon] (o2),
	};
	\end{feynman}
\end{tikzpicture}
  \\[0.7ex]
  -\frac{1}{2}\ii g' \delta^{ij} \gamma_{\mu} P_L
\end{gathered}

&
\begin{gathered}
  \begin{tikzpicture} [baseline=(v.base),scale=0.9,transform shape]
	\begin{feynman}
		\vertex (v);
		\vertex [left=of v] (i1) {\(\overline{L}^{i}_{B}\)};
		\vertex [right=of v] (o1) {\(\overline{L}^{j}_{B}\)};
		\vertex [above=of o1] (o2) {\(B\)};
		\vertex [below=0.25em of v] {\(\mu\)};
		
		\diagram* {
			(i1) -- [anti fermion] (v) -- [anti fermion] (o1),
			(v) -- [photon] (o2),
		};
	\end{feynman}
\end{tikzpicture}
  \\[0.7ex]
  \frac{1}{2}\ii g' \delta^{ij} \gamma_{\mu} P_R
\end{gathered}

&
\begin{gathered}
 \begin{tikzpicture} [baseline=(v.base),scale=0.9,transform shape]
	\begin{feynman}
	\vertex (v);
	\vertex [left=of v] (i1) {\(\Phi^{i}\)};
	\vertex [right=of v] (o1) {\(\Phi^{j}\)};
	\vertex [above=of o1] (o2) {\(B\)};
	\vertex [below=0.25em of v] {\(\mu\)};
	
	\diagram* {
		(i1) -- [charged scalar,momentum'={$p$}] (v) -- [charged scalar,momentum'={$q$}] (o1),
		(v) -- [photon] (o2),
	};
	\end{feynman}
\end{tikzpicture}
  \\[0.7ex]
  \frac{1}{2}\ii g' \delta^{ij}(p+q)_\mu
\end{gathered}

\end{array}
\end{equation*}
Scalar and fermion propagators are denoted in amplitudes as
\begin{eqnarray}
	\ii S_{j}(p) &=& \frac{\ii (\s{p}+m_{j})}{p^{2}-m_{j}^{2}+\ii \epsilon}\;,\\
	\ii D_{j}(p) &=& \frac{\ii}{p^{2}-m_{j}^{2}+\ii\epsilon}\;,
\end{eqnarray}
where propagators for vector bosons are given by $-\ii g_{\mu\nu} D_j(p)$.

\section{Loop Diagrams}\label{app:loops}

In this appendix, we collect the remaining loop diagrams that (together with Eq.~\ref{img:L1}) interfere with the tree-level process. The tree-level process has no complex phases, so a complex phase (e.g., $A_{12}$) is required in the diagram to produce CP-violation. The leading-order contribution arises from diagrams with both a Higgs and a $W$ (or $B$) gauge boson exchange. The cuts that contribute to the computation of $\epsilon$ are drawn with a dashed red line.

\beq\label{img:L2}
\begin{tikzpicture} [baseline=(m1.base)]
	\begin{feynman}
	\vertex (i1) {$\chi_1$};
	\vertex [below=of i1] (m1);
	\vertex [right=of m1] (m2);
	\vertex [right=of m2] (m3);
	\vertex [right=of m3] (m4);
	\vertex [right=of m4] (m5);
	\vertex [right=of m5] (m6);
	\vertex [right=of m6] (m7) {$\Phi^{*m}$};
	\vertex [below=of m1] (i2) {$\chi_1$};
	\vertex [right=of i1] (u1);
	\vertex [right=of u1] (u2);
	\vertex [right=of u2] (u3);
	\vertex [right=of u3] (u4);
	\vertex [right=of u4] (u5);
	\vertex [right=of u5] (o1) {$L_1^{o}$};
	\vertex [above=of u4] (h1);
	\vertex [above=of u5] (h2) {$\Phi^{*n}$};
	\vertex [right=of i2] (b1);
	\vertex [right=of b1] (b2);
	\vertex [right=of b2] (b3);
	\vertex [right=of b3] (b4);
	\vertex [right=of b4] (b5);
	\vertex [right=of b5] (o2) {$L_1^{p}$};
		
	\diagram* {
		(i1) -- [solid] (u1) -- [anti fermion, edge label={$\overline{L}_1^{i}$}] (u3) -- [solid, edge label={$\chi_2$}] (u4) -- [fermion] (o1),
		(u3) -- [anti charged scalar, edge label={$\Phi^{*\ell}$}] (m5) -- [anti charged scalar] (m7),
		(u4) -- [anti charged scalar] (h2),
		(i2) -- [solid] (b1) -- [fermion, edge label'={$L_1^{k}$}] (b4) -- [fermion] (o2),
		(u1) -- [charged scalar, edge label={$\Phi^{j}$}] (b1),
		(b4) -- [photon, edge label={$W^B$}] (m5),
				};
	\draw [dashed, thick, red] (2.5,1) -- (2.5,-4.5);
	\end{feynman}
\end{tikzpicture} 
\eeq

\beq\label{img:L3}
\begin{tikzpicture} [baseline=(m1.base)]
	\begin{feynman}
	\vertex (i1) {$\chi_1$};
	\vertex [below=of i1] (m1);
	\vertex [right=of m1] (m2);
	\vertex [right=of m2] (m3);
	\vertex [right=of m3] (m4);
	\vertex [right=of m4] (m5);
	\vertex [right=of m5] (m6);
	\vertex [right=of m6] (m7) {$\Phi^{*n}$};
	\vertex [below=of m1] (i2) {$\chi_1$};
	\vertex [right=of i1] (u1);
	\vertex [right=of u1] (u2);
	\vertex [right=of u2] (u3);
	\vertex [right=of u3] (u4);
	\vertex [right=of u4] (u5);
	\vertex [right=of u5] (o1) {$L_1^{o}$};
	\vertex [above=of u4] (h1) {$\Phi^{*m}$};
	\vertex [above=of u5] (h2);
	\vertex [right=of i2] (b1);
	\vertex [right=of b1] (b2);
	\vertex [right=of b2] (b3);
	\vertex [right=of b3] (b4);
	\vertex [right=of b4] (b5);
	\vertex [right=of b5] (o2) {$L_1^{p}$};
		
	\diagram* {
		(i1) -- [solid] (u1) -- [anti fermion, edge label={$\overline{L}_1^{i}$}] (u3) -- [solid, edge label={$\chi_2$}] (u4) -- [fermion] (o1),
		(u3) -- [anti charged scalar] (h1),
		(u4) -- [anti charged scalar,edge label'={$\Phi^{*\ell}$}] (m5) -- [anti charged scalar] (m7),
		(i2) -- [solid] (b1) -- [fermion, edge label'={$L_1^{k}$}] (b4) -- [fermion] (o2),
		(u1) -- [charged scalar, edge label={$\Phi^{j}$}] (b1),
		(b4) -- [photon, edge label={$W^B$}] (m5),
				};
	\draw [dashed, thick, red] (2.5,1) -- (2.5,-4.5);
	\end{feynman}
\end{tikzpicture}
\eeq

\beq\label{img:L4}
\begin{tikzpicture} [baseline=(m1.base)]
	\begin{feynman}
	\vertex (i1) {$\chi_1$};
	\vertex [below=of i1] (m1);
	\vertex [right=of m1] (m2);
	\vertex [right=of m2] (m3);
	\vertex [right=of m3] (m4);
	\vertex [right=of m4] (m5);
	\vertex [right=of m5] (m6);
	\vertex [right=of m6] (m7);
	\vertex [below=of m1] (i2) {$\chi_1$};
	\vertex [right=of i1] (u1);
	\vertex [right=of u1] (u2);
	\vertex [right=of u2] (u3);
	\vertex [right=of u3] (u4);
	\vertex [right=of u4] (u5);
	\vertex [right=of u5] (o1) {$L_1^{o}$};
	\vertex [above=of u4] (h1) {$\Phi^{*m}$};
	\vertex [above=of u5] (h2) {$\Phi^{*n}$};
	\vertex [right=of i2] (b1);
	\vertex [right=of b1] (b2);
	\vertex [right=of b2] (b3);
	\vertex [right=of b3] (b4);
	\vertex [right=of b4] (b5);
	\vertex [right=of b5] (o2) {$L_1^{p}$};
		
	\diagram* {
		(i1) -- [solid] (u1) -- [anti fermion, edge label={$\overline{L}_1^{i}$}] (u3) -- [solid, edge label'={$\chi_2$}] (u4) -- [fermion, edge label'={$L_1^{\ell}$}] (u5) -- [fermion] (o1),
		(u3) -- [anti charged scalar] (h1),
		(u4) -- [anti charged scalar] (h2),
		(i2) -- [solid] (b1) -- [fermion, edge label'={$L_1^{k}$}] (b5) -- [fermion] (o2),
		(u1) -- [charged scalar, edge label={$\Phi^{j}$}] (b1),
		(u5) -- [photon, edge label={$W^B$}] (b5),
				};
	\draw [dashed, thick, red] (2.5,1) -- (2.5,-4.5);
	\end{feynman}
\end{tikzpicture}
\eeq

\bibliographystyle{JHEP}
\bibliography{references}

\end{document}